\documentclass[aps,showkeywords,groupedaddress,nofootinbib]{revtex4}
\usepackage{graphicx}
\usepackage{color}
\usepackage{amsmath}
\usepackage{amsfonts}
\usepackage[toc]{appendix}

\usepackage{lineno}



\begin{document}

\title{RBE with Non-Poisson Distribution of Radiation Induced Strand Breaks}

\author{M. Loan$^{a,b}$\footnote{Corresponding author}, M. Alameen$^{c}$, A. Bhat$^{d}$ and M.
Tantary$^{e}$}
\affiliation{$^{a}$Department of Physics, Kuwait College of Science and Technology, 28007, Kuwait \\
$^{b}$ANUC, Australian National University, Canberra, 2000, Australia\\
$^{c}$Department of Mathematics, Australian College of Kuwait, 13015, Kuwait\\
$^{d}$Department of Radiation Oncology, Clinch Valley Medical Center, Richlands, Virginia, 24641, USA\\
$^{e}$Department of Internal Medicine, Clinch Valley Medical
Center, Richlands, Virginia, 24641, USA}
\date{\today}

\begin{abstract}

\noindent{Postulating that increasing linear energy transfer (LET)
causes non-random clustering of lethal lesions to deviate from the
Poisson distribution, we employ a non-Poisson approach as a more
flexible alternative that accounts for overdispersion of lethal
lesions. Using non-homologous end-joining (NHEJ) pathway of
double-strand break repair, a customized negative binomial (NB)
distribution is used to describe the distribution of lethal events
in a cell nucleus. The proposed model provides a novel,
mechanistically based explanation for the measured values of the
biological relevant quantities, such as model parameters and
relative biological effectiveness (RBE) of the surviving cells,
for various light ion types and LET values.  The estimated
quantities are compared with the predictions of several
mechanism-inspired models and experimental data at medium and high
LET values. The results examined are closer to the
Microdosimetric-Kinetic model predictions for helium and carbon
ions but progressively lower than trends predicted by the Local
Effect Model and the Repair-Misrepair-Fixation model in the large
LET region. The results support the view that the limitation in
the increase in RBE at high LET can be accounted for entirely, or,
in large part, by clustering of lethal events to cause deviation
from the Poisson distribution.}\\


\end{abstract}

\maketitle

\section{Introduction}

It has long been established  that the linear-quadratic (LQ) model
is a good approximation to a wide range of damage-kinetic models
which describe the kinetics of DNA double-strand breaks (DSBs) and
other basic lesions
\cite{Guerrero2002,Preston1990,Sachs1997,Gerweck1994,Thames1985,Brown2003,Stewart2001}.
Underlying the application of these kinetic-models to
fractionation effects is pairwise misrepair of primary lesions
such as DSBs or base damage. The double-strand breaks are resolved
either through restitution or binary misrepair. At typical
radiotherapeutic doses, most DSBs are removed by restitution,
which results in the classic linear-quadratic dose dependence. At
very high doses per fraction, binary misrepair can dominate, which
results in a linear relation between effect and dose. Overall,
these mechanisms produce a linear-quadratic-linear dose-response
relationship, as has been pointed out by many authors
\cite{Rossi1988,Brenner1990,Radivoy1998,Carlone2005}. Detailed
experimental studies in vivo cell survival suggested that all
these data  are consistent with the Linear-Quadratic (LQ) model up
to about 20 - 24 Gy
\cite{Barendsen1997,vander1985,Peck1994,Taylor1989,de1988}. In
vitro cell survival, the quality of fit to the LQ model does not
decline significantly until doses above 15 Gy are included
\cite{Garcia2006}. Theoretical estimates, on the other hand,
gauged the practical applicability of LQ approximation at dose
below 17 Gy and suggested corrections to the LQ model at higher
doses \cite{Sachs1997}.  Various other mechanistic models
describing pairwise production of chromosome aberrations predict
virtually the same time-dose relations as does the LQ approach
\cite{Brenner1998,Curtis1986,Hawkins1996,Obaturov1993,Tobias1995}.

Although different models have different formulations for the dose
dependence of the mean lethal lesion yield, the common feature of
the approaches used in these models is that error distribution
around the mean number of lethal lesions per cell is assumed to be
Poisson. Such assumptions are understandable, given that these
models evolved to explain clonogenic cell survival, an endpoint
for which the number of lesions in individual cells cannot be
quantified. Additionally, the linear relationship of RBE and LET
is consistent with the experimental values for LETs low enough
that the radiation induced lethal lesions are Poisson distributed
among the cells of the irradiated population
\cite{Hawkins1996,Hawkins1998}. This linearity allows
extrapolation to values of LET greater than the range for which
there is Poisson distribution of lethal lesions. However, despite
widespread usage of the LQ, there remain questions about departure
of its linear quadratic relation from the experimental survival
values \cite{Zaider1998}. The experimentally measured RBE with
increasing LET becomes progressively less than that of the linear
increase established for LET $< 70$ keV/$\mu$m \cite{Hawkins2017}.
This could be due to the deviation of the distribution of lethal
lesions among the irradiated cells from that of the Poisson
distribution.  Whereas for sparsely-ionizing radiation below $\sim
50$ Gy per fraction, the deviations from a Poisson distribution
are quite minor \cite{Sachs1997,Sachs1998}, the deviations of
lethal lesions from Poisson distribution are significant at high
LETs \cite{Zaider1998,Brenner2008,Kirk2009,Hawkins2003}. This is
evident from chemical and biological factors, such as
heterogeneity of DSB complexities, the existence of multiple DSB
repair pathways with different fidelities as well as from the
microdosimetric considerations for ionizing radiations such as
hadrons and heavy ions at higher LETs and doses
\cite{Harrison2014,Iliakins2015,Shelke2015,Hawkins2017}.  It is
plausible to hypothesize that these processes can lead  to
overdispersion which can change the predicted cell survival curve
shape.

As a result, many investigators have raised questions about the
general validity of the LQ, and whether it truly represents
underlying biology or is simply a useful empirical tool
\cite{Hawkins2003,Kirk2009,Hawkins2017b}. The discrepancy between
experimental survival and the linear quadratic model at higher
dose is explained by heterogeneity of sensitivity to radiation
among the cells of the irradiated population \cite{Hawkins2017b}.
Reported increase in local multiplicity of radiation-induced
strand-breaks with increasing LET could constitute such a change
in lesion structure \cite{Ward1988,Goodhead1994} as well as a
decrease in DSB yield for LET $> 100$ keV/$\mu$m
\cite{Heilmann1995}. In addition to the shift in  maximum  of RBE,
the resulting RBE is estimated to be less than indicated by
extrapolation of  the linear relationship to higher LET values
\cite{Hawkins2003,Brenner2008}. Postulating that increasing LET
causes non-random clustering of lethal lesions in some cells to
deviate from the Poisson distribution, we use a non-Poisson
distribution as a more flexible alternative that allows to
accommodate a variety of mechanisms for overdispersion. Other
non-Poisson distributions used to target overdispersion that
affects the predicted dose response for cell survival include, for
example, the Neyman distribution, which accounts for stochasticity
of the number of ionizing track traversals per cell and the number
of chromosomal aberrations per track
\cite{Nowak2000,Virsik1981,Goodwin1994}. We employ a customized
negative binomial (NB) distribution to derive an equation that
predicts the model parameters which are used to estimate RBE of
V79 cell lines radiated by charged hadrons with different physical
parameters. The dependence of the predicted RBE on LET is compared
with the results obtained from the Local Effect Model (LEM)
\cite{Elsasser2010,Friedrich2012}, the Microdosimetric-Kinetic
(MK) Model \cite{Hawkins1994,Inaniwa2010}, and the
Repair-Misrepair-Fixation (RMF) Model
\cite{Carlson2008,Frese2012,Streitmatter2017}. The estimated RBE
behaviour, in principle, may offer a clinically useful approach
for modelling the effects of high doses per fraction at large LET.

\section{Non-Poisson Distribution of Lethal Lesions}
\subsection{Model and Method}
Assuming the lethal lesions are Poisson distributed from cell to
cell, the surviving probability of a cell is given by
\begin{equation}
S = e^{-Y} = e^{-\alpha D - G\beta D^{2}},
\label{eqn1}
\end{equation}
where $Y$ is the yield of lethal lesions, $\alpha$ and $\beta$ are
parameters describing the cell's radiosensitivity, and $G$ is the
generalized Lea-Catcheside time factor, which accounts
quantitatively for fractionation/protraction. It has been well
established that the effective plot of Eq. (\ref{eqn1}), on a log
scale, gives what is referred to as a ''shouldered'' dose response
curve. The initial region of the curve is dominated by the linear
term at low doses, followed by increasing curvature as the
quadratic term becomes more significant. The degree of curvature
is commonly expressed in terms of the $\alpha / \beta$ ratio or
cellular repair capacity (CRC) and corresponds to the dose at
which the linear and quadratic contributions are equal. Thus,
cells with high CRC  see a relatively constant rate of cell
killing with increasing dose, while those with a low CRC show a
pronounced curvature. Assuming that there is no reduction in cell
killing due to repair ($G=1$), the simplest single-fraction LQ
formalism takes the form:
\begin{displaymath}
S = e^{-\alpha D - \beta D^{2}}.
\end{displaymath}
Also, assuming that the distribution of lethal chromosomal lesions
among individual human cells exposed to light ion irradiation is
somewhat overdispersed,  we employ a customized negative binomial
(NB) distribution for evaluating the probability of observing $k$
lethal lesions in a cell
\begin{equation}
P_{NB}(k) = \frac{\Gamma (k+1/r)}{\Gamma (1/r)\times k!}
\bigg(\frac{1}{1+r\mu }\bigg)^{1/r} \times
\bigg(\frac{1}{1+1/r\mu}\bigg)^{k}.
 \label{eqn2}
\end{equation}
Specifying the distribution in terms of its mean, $\mu=\lambda =
(1-p)r/p$, the variance is described by the convenient expression
$\mu +\mu^{2}/r = \mu+\omega \mu^{2}$, where either the parameter
$r$ or its reciprocal $\omega$ is referred to as the
over-dispersion parameter. Consequently, if $\omega \rightarrow
0$, there is no over-dispersion and the variance and mean are
equal, as in the Poisson distribution. On the other hand, if
$\omega >0$, the variance becomes greater than the mean and the
ratio of variance to mean increases as the mean increases. The
same approach is extended to model the mean yield $Y$ of lethal
lesions per cell after a radiation dose $D$, with the cell
surviving fraction $S$ defined as the probability of zero lethal
lesions as:
\begin{equation}
S_{IP} =P_{NB}(0)= (1+r Y)^{-1/r}.
\label{eqn3}
\end{equation}

Following \cite{Wang2018}, the radiation-induced DSBs in the
nucleus and probability of cell death can be predicted by first
calculating the average number of primary particles that cause
DSB, $n_{p}$, and the average number of DSBs yielded by each
primary particle that causes DSB, $\lambda_{p}$. The DSB yield per
cell per primary particle is given by $\lambda = N/n$, where
$N=YD$ is the average number of radiation-induced DSBs per cell
and $n$ is the number of the particles passing through the cell
nucleus. Assuming that the number of DSBs yielded by a primary
particle is distributed according to NB distribution, Eq.
(\ref{eqn2}), the probability of a primary particle passing
through a nucleus without causing any DSB is given by
\begin{equation}
P_{NB}(k=0)= (1+r \lambda)^{-1/r}.
\label{eqn4}
\end{equation}
The average number of primary particles that cause DSB, and the
average number of DSBs yield per primary particle that causes DSB
are respectively given by
\begin{eqnarray}
n_{p} & = & n[1-(1+r \lambda )^{-1/r}],\\
\lambda_{p} & = & \frac{\lambda}{[1-(1+r \lambda )^{-1/r}]}.
\label{eqn5}
\end{eqnarray}

The repair of the breaks is modelled by the non-homologous
end-joining (NHEJ) pathway \cite{Karge2017} with the overall
behaviour given by
\begin{equation}
N_{avg} =\mu_{y}N\times (1-P_{correct}),
\label{eqn6}
\end{equation}
where $N_{avg}$ is the average number of lethal events, $\mu_{y}$
is the sensitivity of an error repair, and $P_{correct}$ is the
total probability of a DSB being correctly repaired. For any break
in a particular condition, the true distribution of rejoining
rates is a non-trivial and complex function and lacks a simple
distribution for its overall distribution. Following
\cite{McMahon2016,Wang2018}, the repair behaviour is approximating
the recombination function as a step function
\begin{displaymath}
f(x) = \left\{ \begin{array}{ccc}
         1 & \mbox{if $x\leq d_{max}$}\\
        0 & \mbox{elsewhere},\end{array} \right.
\end{displaymath}
where $d_{max}$ is some maximum separation between the break ends.
This simplifies the probability of correct repair for a given
break to $1/(1+k )$, where $k$ is the number of breaks within the
distance $d_{max}$ \cite{McMahon2016}. Assuming that the breaks
within the spherical radius are NB distributed, the expectation
value that a randomly chosen break will repair correctly (a DSB
end do not be joined with a DSB end from a DSB induced by a
different primary particle) is given by (see the appendix)
\begin{equation}
P_{1} = \frac{1}{\mu_{int}(1-r)}\left[1- (1+
r\mu_{int})^{1-1/r}\right],
\label{eqn7}
\end{equation}

where $\mu_{int}=\eta (\lambda_{p})n_{p}$ is the average
probability of a DSB end being joined with a DSB end from  a DSB
induced by a different primary particle and $\eta$ is the
effective integral misrejoining probability for a single randomly
chosen DSB within a radius.

Also, assuming that a primary particle generates DSBs randomly on
it's track, the probability that a DSB end not be joined with a
DSB end from a different DSB induced by the same primary particle
is given by
\begin{equation}
P_{2} = \frac{1}{\mu_{track}(1-r)}\left[1- (1+
r\mu_{track})^{1-1/r}\right],
\label{eqn8}
\end{equation}
where $\mu_{track}=\phi\lambda_{p}$ is the average probability of
a DSB end being joined with a DSB end from  a DSB induced by the
same primary particle. Therefore, the total probability of a DSB
being correctly repaired is given by
\begin{eqnarray}
P_{correct} & = &
\mu_{x}P_{1}P_{2}\nonumber\\
& = & \mu_{x}\left[\frac{1}{\mu_{int}(1-r)}\bigg(1- (1+r
\mu_{int})^{1-1/r }\bigg)\right]\times
\left[\frac{1}{\mu_{track}(1-r)}\bigg(1-
(1+r \mu_{track})^{1-1/r}\bigg)\right],\nonumber\\
 \label{eqn9}
\end{eqnarray}
where $\mu_{x}$ is process-specific fidelity. The first term in
the square brackets quantitatively describes the interaction of
DSBs induced by different primary particles and the second term in
square brackets describes the effect of clustered DNA damage.

Using the above repair probabilities, Eq. (\ref{eqn6}) becomes
\begin{eqnarray}
N_{avg} &= & \mu_{y}N\times \left[1-
\mu_{x}\left\{\frac{1}{\mu_{int}(1-r)}\bigg(1- (1+r
\mu_{int})^{1-1/r}\bigg)\right\} \right. \nonumber\\
& & \times \left. \left\{\frac{1}{\mu_{track}(1-r)}\bigg(1- (1+r
\mu_{track})^{1-1/r}\bigg)\right\}\right],
 \label{eqn10}
\end{eqnarray}
The cell survival becomes
\begin{equation}
S_{NB} =  \bigg(1+ rN_{avg}\bigg)^{-1/r}= \bigg(1+r(\alpha D+\beta
D^{2})\bigg)^{-1/r},
\label{eqn11}
\end{equation}
where
\begin{eqnarray}
\alpha & = & \mu_{y}Y\left[1 - \mu_{x}\frac{1}{\phi\lambda_{p}(1-r)}\bigg(1-(1+r\phi\lambda_{p})^{1-1/r}\bigg)\right]\nonumber\\
\beta & = & \frac{1}{2}\frac{\eta
(\lambda_{p})}{\lambda_{p}}\mu_{x}\mu_{y} Y^{2}
\frac{1}{\phi\lambda_{p}(1-r)}\bigg(1-(1+r\phi\lambda_{p})^{1-1/r}\bigg)
 \label{eqn12}
\end{eqnarray}
are the improved parameters of the model. The values of $\mu_{x},
\mu_{y}, \xi, \phi$ and $\eta_{\lambda_{p}\rightarrow 1}$ obtained
with the experimental data of linear parameter values and
$\eta_{\lambda_{p}\rightarrow \infty}$ with the experimental data
of $\beta$ values of the survival curves are adapted from Wang et
al \cite{Wang2018}.\\

The yield of DSBs induced by ionizing radiations was calculated
with fast Monte Carlo damage simulation (MCDS) software
\cite{MCDScode}. This algorithm captures the trend in DNA damage
spectrum with the possibility that the small-scale spatial
distribution of elementary damages is governed by stochastic
events and processes \cite{Stewart2015,Stewart2018}. The use of
this quasi-phenomenological algorithm is to provides
nucleotide-level maps of the clustered DNA lesions and to avoid
the initial simulation of the chemical processes. It has been
observed that MCDS algorithm gives reliable results of the damage
yields that are comparable to those obtained from computationally
expensive but more detailed track structure simulations. For MCDS
simulations, the results of DNA damage yields for protons and
light ions are usually obtained within minutes. We generate ten
ensembles of DSB yield measurements for each LET value for later
analysis. The expectation values and statistical error estimates
of the observables  were estimated using the jackknife method. The
statistical errors were estimated by grouping the stored
measurements into 5 bins, and then the mean and standard deviation
of the final quantities were estimated by averaging over the bin
averages. The estimate of the error of observable $\rho$ was
calculating by using
\begin{displaymath}
\delta \rho = \sqrt{\frac{M-1}{M}\sum_{m=1}^{M}(\bar{\rho}_{m}
-\langle \rho \rangle )^{2}}.
\end{displaymath}
Statistical significance between two data sets was accessed using
a 2-tail $t$-test with a $95\%$ confidence interval.\\

Using the maximum likelihood estimation to fit the negative
Binomial distributions to the survival data on 1.15 MeV proton
irradiated V79 cells and 25 MeV helium ion irradiated T1 cells
\cite{Prise1990}, we found parameter values that maximize the
likelihood of making the observations given the parameters. A
log-likelihood maximization (Appendix Eq. A2) was performed across
the analyzed cells to find the best-fit value for a particular
data set. The NB log-likelihood function was performed using the
parmhat function that specifies control parameters for the
iterative algorithm the function uses. The programming was
implemented in MATLAB $\copyright$R2018a software. At $95\%$
confidence intervals, the value of $r$ was found  to be 0.043 and
0.241, for the lower and upper limits, respectively. The negative
binomial overdispersion parameter was set to 0.142, the best-fit
value for data on V79 cell survival.

\subsection{Particle $RBE$ Characterization}
Whereas a large amount of data from different experimental
protocols and biological models are available
\cite{Friedrich2013}, the adoption of a simple and unique RBE-LET
relationship in effective treatment planning is surrounded by a
number of uncertainties. Few studies have supported a reasonable
approximation of fixed RBE to describe the increased effectiveness
of light ions
\cite{Paganetti2002,Paganetti2014,Giovannini2016,Leeuwen2018}, the
concerns for a better understanding of RBE-LET relationship for
significant clinical relationship have been raised. The modelled
parameters $\alpha$ and $\beta$ for the same type of cells
irradiated by different radiation types at different LET can be
used to reflect on the relative biological effectiveness (RBE).
The RBE for cell killing by high LET radiation is conventionally
defined in terms of the ratio of the low LET reference radiation,
$(\alpha/\beta)_{x}$, to the high LET dose producing the same
survival fraction. However, it is convenient to focus on the RBE
in terms of its asymptotic values. Following \cite{Mario2017}, we
consider the effect of changed production of sub-lethal damage
with varying LET by incorporate the twin concepts of $RBE_{max}$
and $RBE_{min}$, which represent the asymptotic values  of RBE in
the limit of $0$ and $\infty$ dose, respectively. In these limits,
\begin{displaymath}
RBE_{max} = \alpha/\alpha_{x}, \hspace{1.0cm} RBE_{min} =
\sqrt{\beta/\beta_{x}},
\end{displaymath}
where $\alpha$ and $\beta$ are given by Eq.(\ref{eqn12}). The
resulting expression for RBE is given by \cite{Mario2017}
\begin{equation}
RBE = (-(\alpha /\beta )_{x}+\sqrt{\Gamma})/2D_{ion},
\label{eqn13}
\end{equation}
where
\begin{eqnarray}
\Gamma &= & (\alpha /\beta )^{2}_{x}RBE^{2}_{max}+4(\alpha /\beta
)_{x}RBE^{2}_{max}D_{ion} +4RBE^{2}_{min}D^{2}_{ion}. \nonumber
\end{eqnarray}
We expect the over-dispersion of lethal lesions to affect the
predicted RBE at high LET and doses.

\section{Results and Discussion}
\subsection{Trends in Radiosensitivity with Particle LET}

\begin{figure}[!ht]
\begin{tabular}{ccccc}
\scalebox{0.45}{\includegraphics{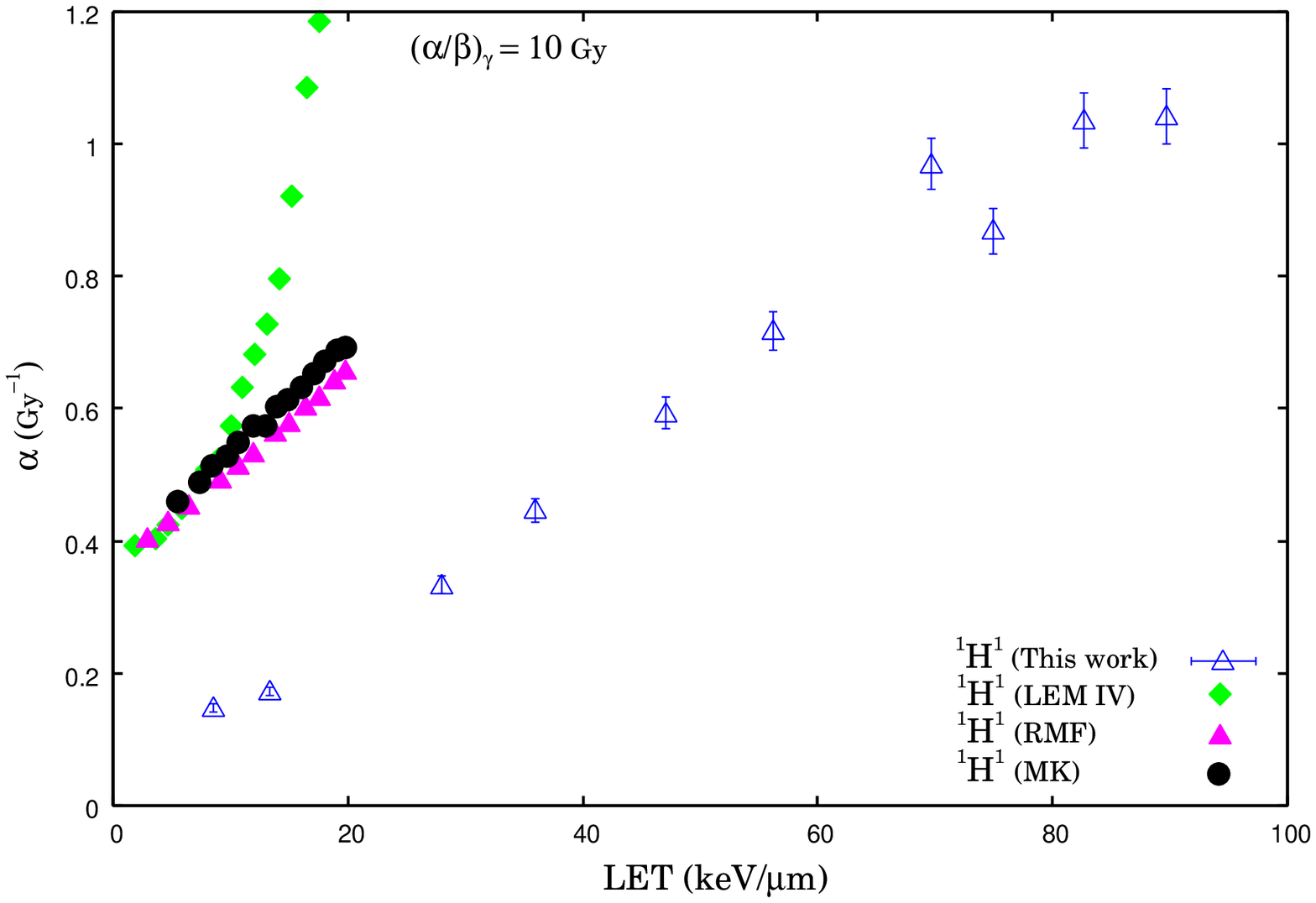}} &
\scalebox{0.45}{\includegraphics{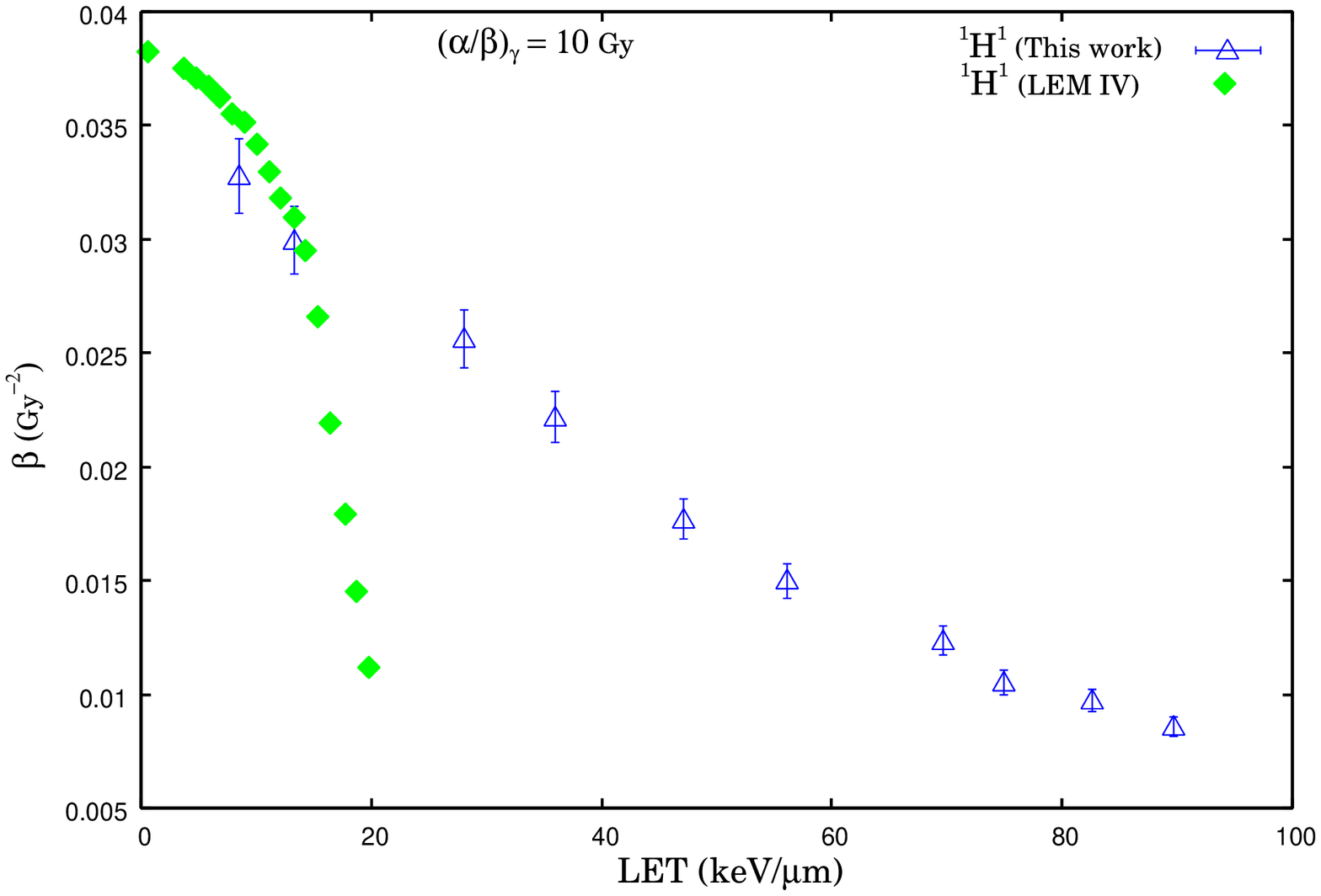}}\\
\scalebox{0.45}{\includegraphics{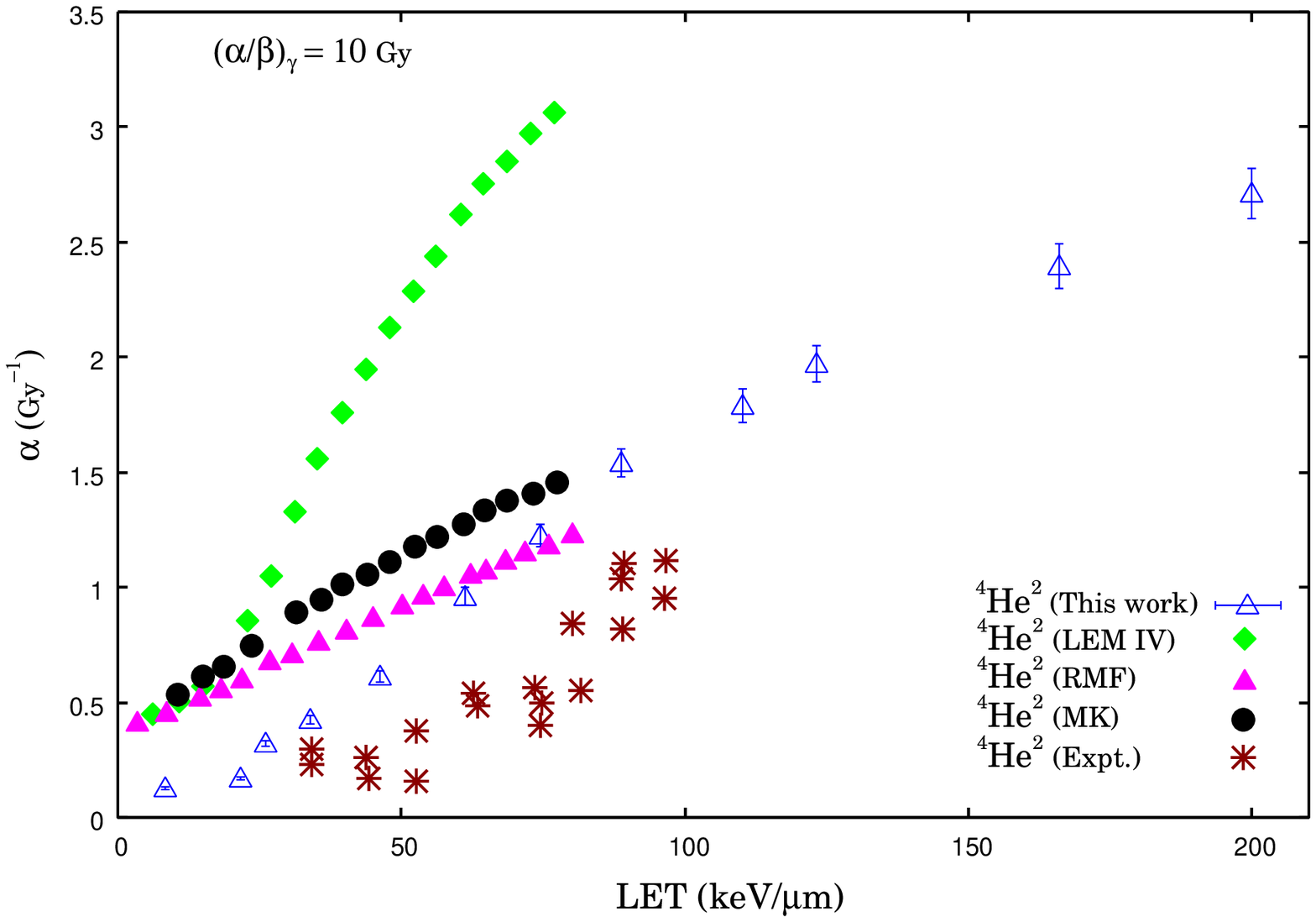}}&
\scalebox{0.45}{\includegraphics{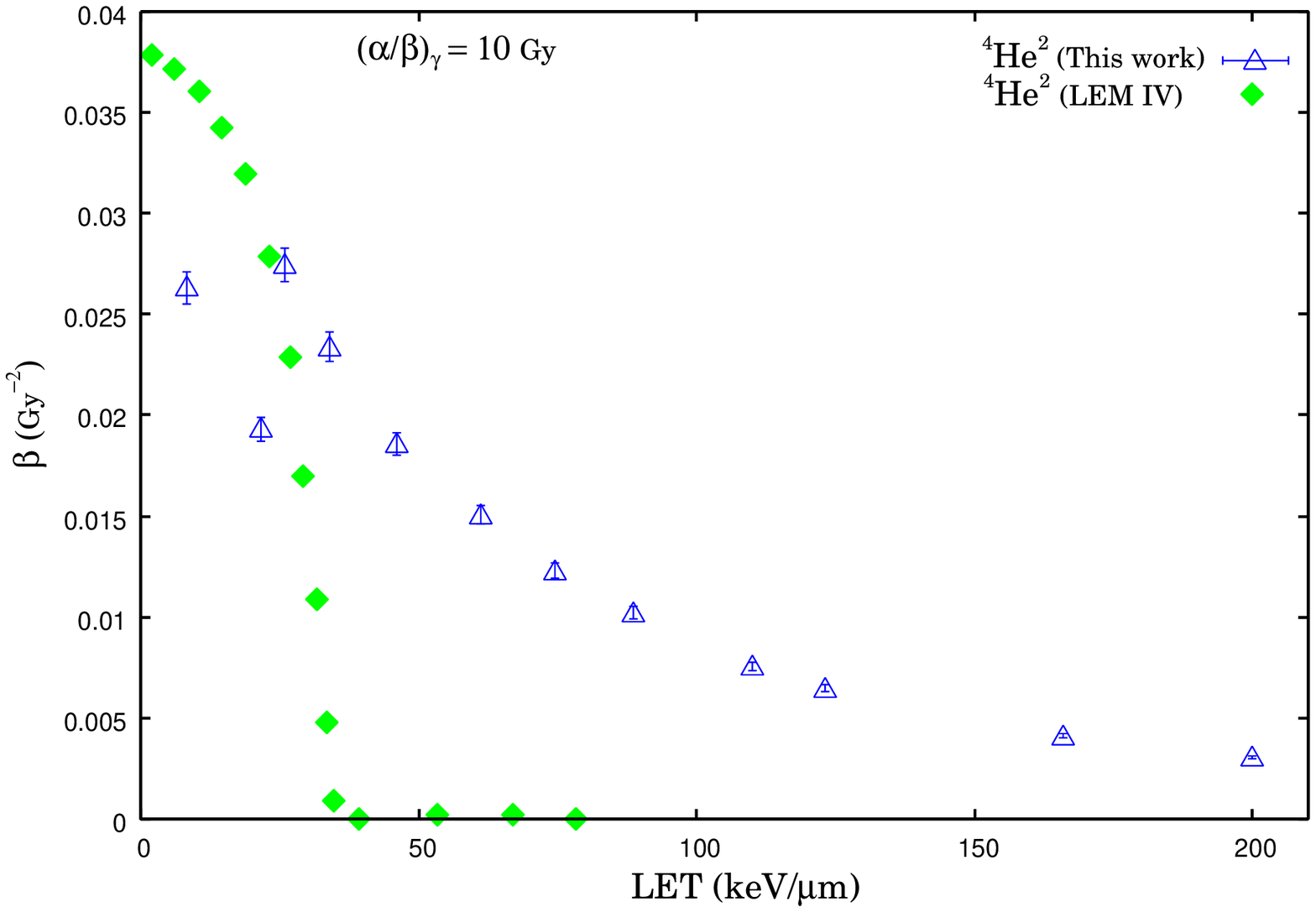}}\\
\scalebox{0.45}{\includegraphics{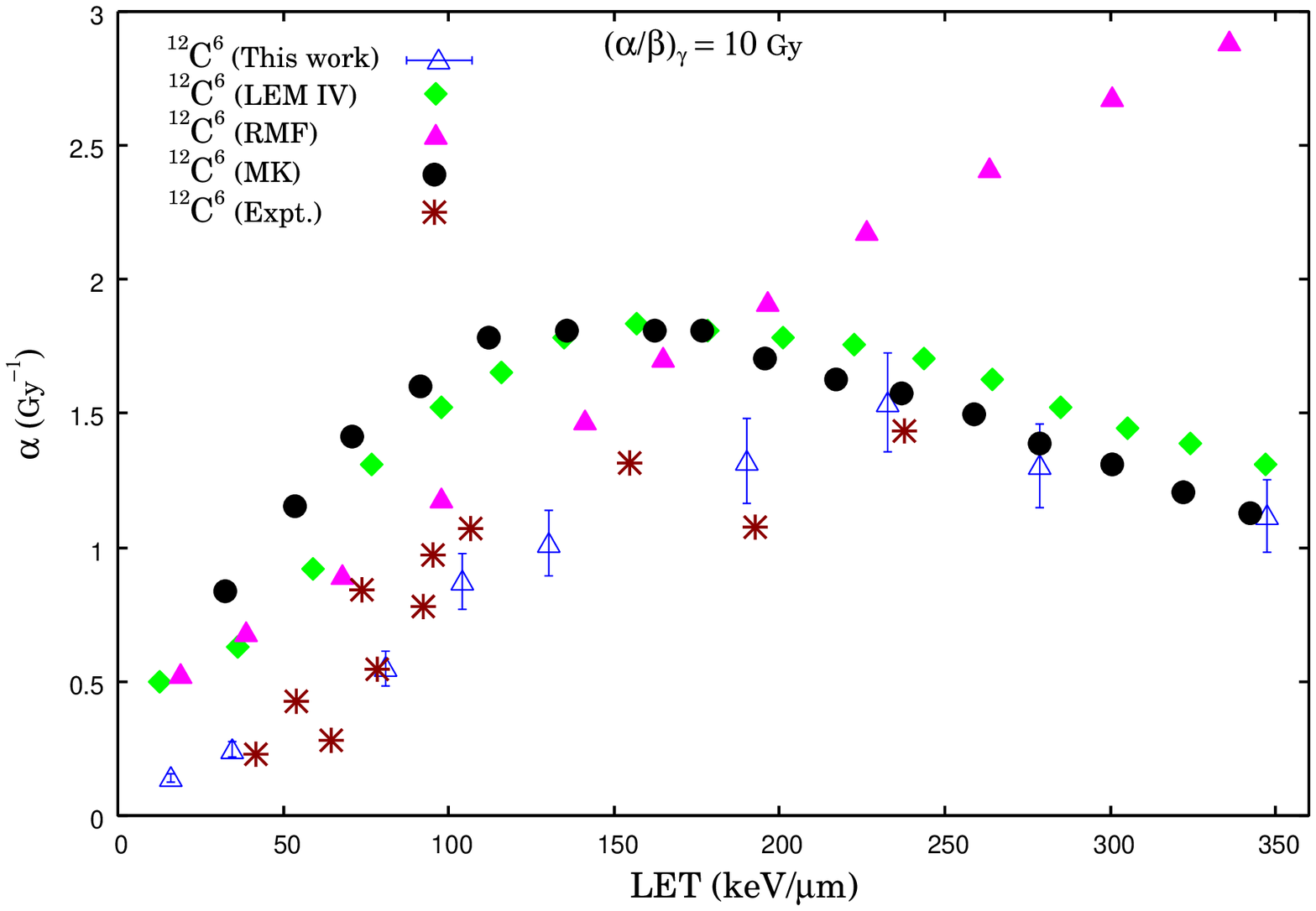}}&
\scalebox{0.45}{\includegraphics{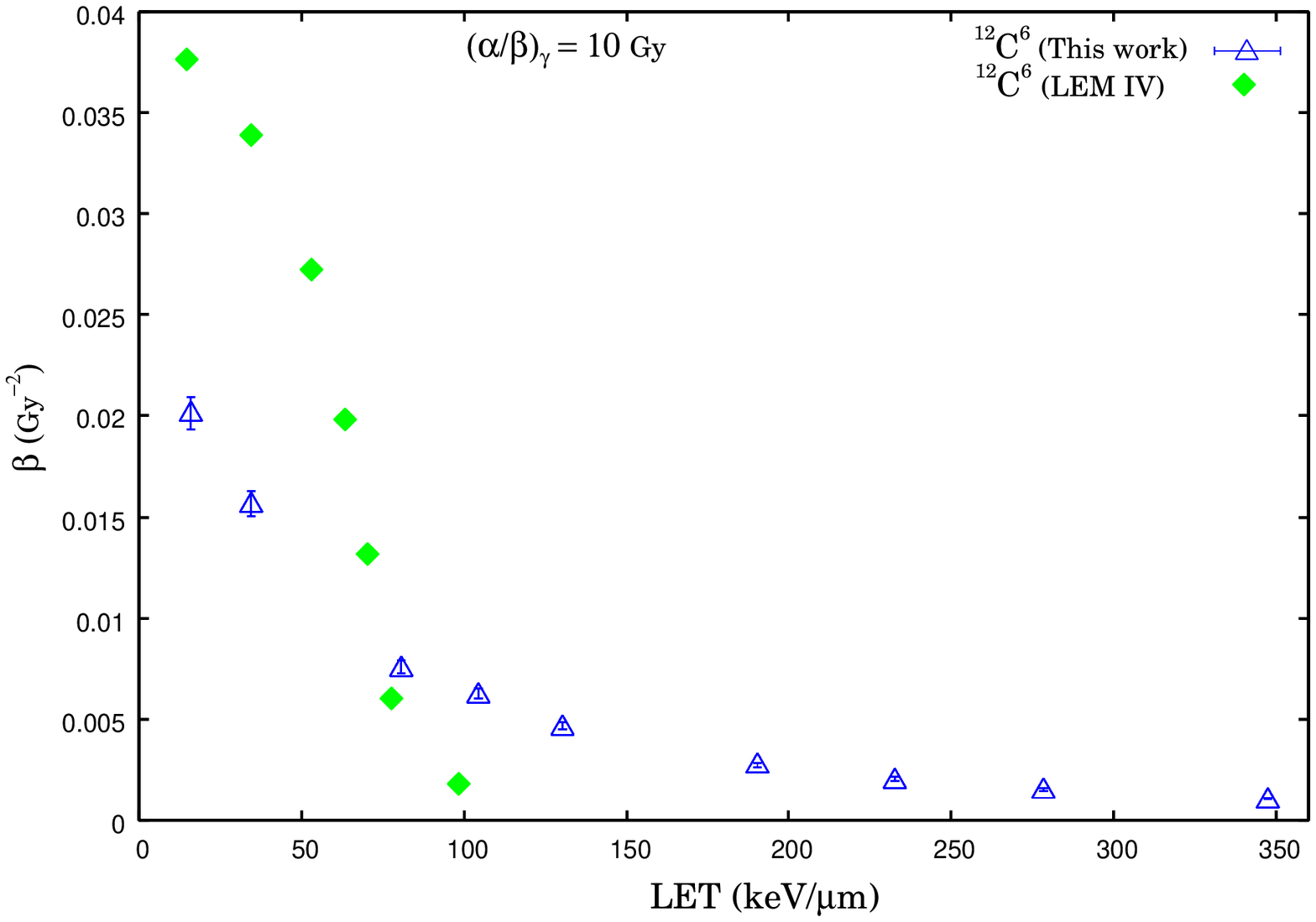}}
\end{tabular}
\caption{\label{fig1} Distribution of linear-quadratic parameters
of the model exposed to protons, helium and carbon ions for a
range of LET values. The experimental results of the linear
parameter using He-3 and carbon ions for a V79 cell
\cite{Furusawa2000} are shown with solid circles. For comparison,
the predictions of the LEM IV, RMF, and MK models
\cite{Stewart2018b} are also shown.}
\end{figure}

Figure \ref{fig1} shows the estimated radiosensitivity parameters
using protons, helium and carbon ions over a wide range of LET
values with the reference radiosensitivity parameters $\alpha_{x}
= 0.38$ $\mbox{Gy}^{-1}$ and $\beta_{x}= 0.038$ $\mbox{Gy}^{-2}$
relative to $^{60}\mbox{Co}$ $\gamma$ rays.. The general trends in
the linear parameter are similar between the proton and
helium-ions, with the measured parameters reaching their
respective maximum between $80$ keV/$\mu$m and $200$ keV/$\mu$m.
At higher LET, the $\alpha$ parameter for helium ions is higher
than that for the protons and carbon ions.  The linear parameter
for carbon ions seems to increase to a maximum before starting to
fall, with the fall-off shifting to higher LET. In terms of the
absolute values, we observe significant variations in our linear
estimates at low LET values. This may be due to the non-Poisson
distribution based description of the model parameters within our
framework.

A comparison of the predicted linear parameter results for helium
and carbon ions with measured data published by Furusawa et al.
\cite{Furusawa2000} shows an underestimation for helium ions for
LET $> 50$ keV/$\mu$m but a reasonable agreement for carbon ions
in both low and high LET regions. Comparison with the LEM IV, RMF,
and MK models \cite{Stewart2018b} shows that our estimates for
protons deviate significantly from the predictions of the these
model, which predict substantially large $\alpha$ values with a
general trend towards a monotonic increase in $\alpha$ values with
increasing LET. For helium ions, our results are comparable,
within errors, with the RMF model and show a comparable trend with
MK model for LET $> 60$ keV/$\mu$m. Our $\alpha$ values for carbon
ions seem to agree with those predicted by MK model for LET $>
200$ keV/$\mu$m. The near agreement with MK model predictions is
not surprising since the distribution of intial DSB in MK model is
effectively formulated as a non-Poisson distribution. The downward
trends in $\alpha$ within the framework of LEM IV model is thought
to be due to the proximity effect for DSB clustering
\cite{Butkus2018}.

The quadratic parameter values suggest a decreasing trend with
increasing LET for all ion species explored here. The predicted
trend is in contrast with the LEM IV model which predicts a sharp
drop in $\beta$ value that eventually approaches zero with
increasing LET. Such a trend is attributed to the simultaneous
decrease in the track diameter with decreasing energy and the
increase in the inter-track distance among DSB with increasing LET
\cite{Stewart2018b}. The MK and RMF model predictions are not
included for comparison since in MK model $\beta$ is considered to
be LET independent and RMF predicts a continuous increase in
$\beta$ with LET which is in contrast with the model analysis of
the data from the Particle Irradiation Data Ensemble (PIDE)
database that suggests  a decreasing trend of $\beta$ with
increasing LET. However, the quantitative analysis of such trends
is needed to draw a definite conclusion.\\

It is obvious from the effective plots that LET strongly affects
linear coefficient than quadratic parameter. It is now generally
accepted that high-LET radiation induces complex clustered damage
to DNA, particularly complex DSBs
\cite{Goodhead1994b,Hada2008,Goodhead2009,David2014}, which are
less repairable by either non-homologous end-joining or the
homologous recombination pathway \cite{Hunnagl2015}. If complex
DSBs are more persistent and more likely to be misrepaired, one
might expect the beta term to increase as the LET increases.

\begin{figure}[!ht]
\begin{center}
\begin{tabular}{ccccc}
\scalebox{0.45}{\includegraphics{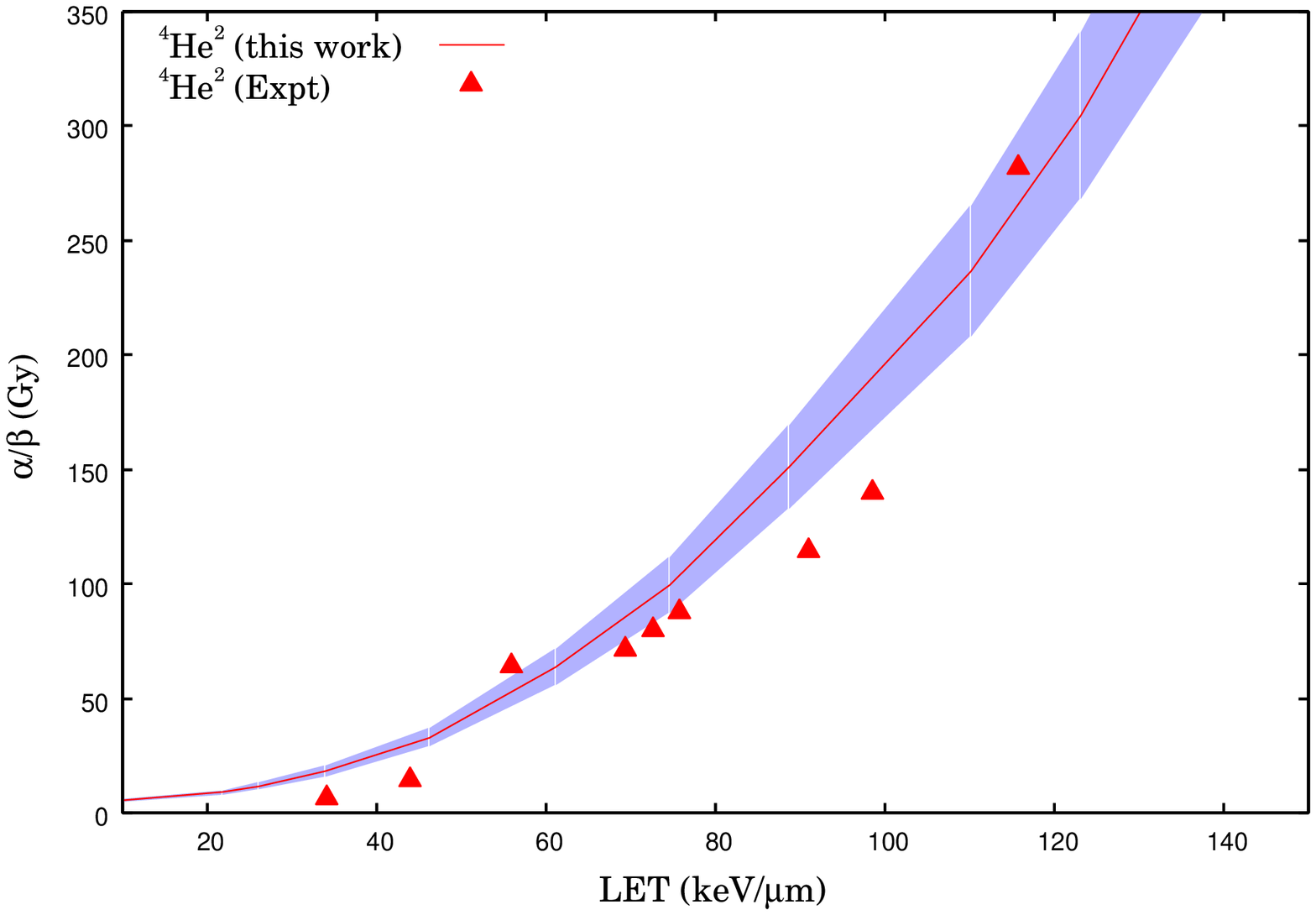}} &
\scalebox{0.45}{\includegraphics{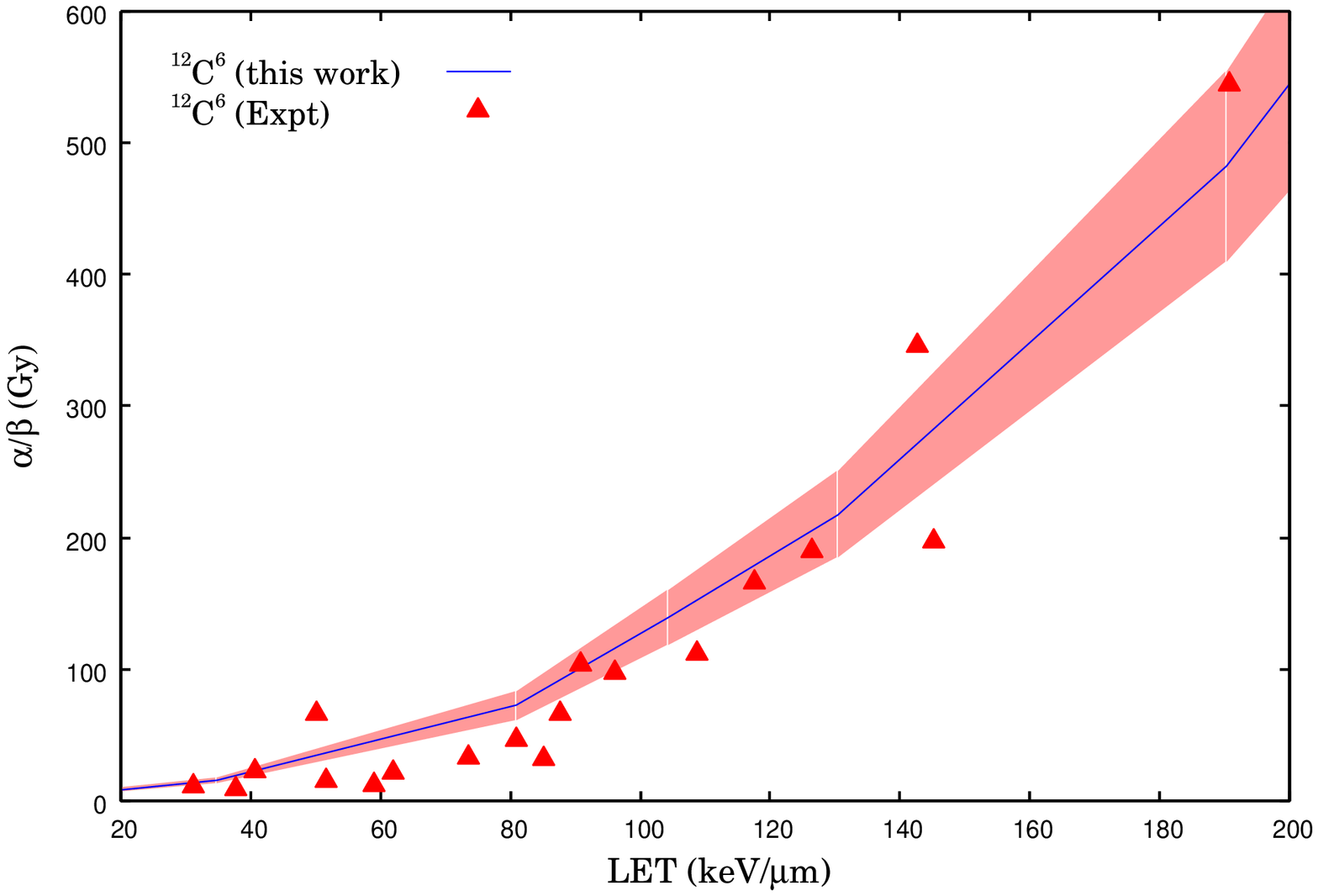}}
\end{tabular}
\caption{\label{fig2} Comparison of modelled results (solid line)
and experimental data (solid triangles) \cite{Furusawa2000} of
$\alpha /\beta$ ratio for V79 cells for helium and carbon ions.}
\end{center}
\end{figure}

To explore the effect of the cell sensitivity on cell killing for
charged particle species, we study the ratio $\alpha /\beta$ which
is mainly attributed to the number of primary particles that cause
DSBs per dose for charged particles at given LET values. Fig.
\ref{fig2} shows the ratio $\alpha /\beta$ for V79 cells
irradiated by helium and carbon ions at different LET along with
the experimental data from \cite{Furusawa2000}. We observe that
the slope of the ratio increase steeply with increasing LET. As
mentioned above, this could be due to increase in $\alpha$ due to
the clustered DNA damage effect as well as the decrease in the
interaction of DSBs induced by different primary particles. This
interaction becomes vanishingly small at intermediate LET values.
We also notice that the ratio tends to reach more or less a
plateau (not shown in the figure) for carbon ions at LET $> 280$
keV/$\mu$m. This might be due to the saturation in the clustered
DNA damage and the effect of overkill on cell death.

\subsection{RBE Results}

\begin{figure}[!ht]
\begin{center}
\begin{tabular}{ccccc}
\scalebox{0.45}{\includegraphics{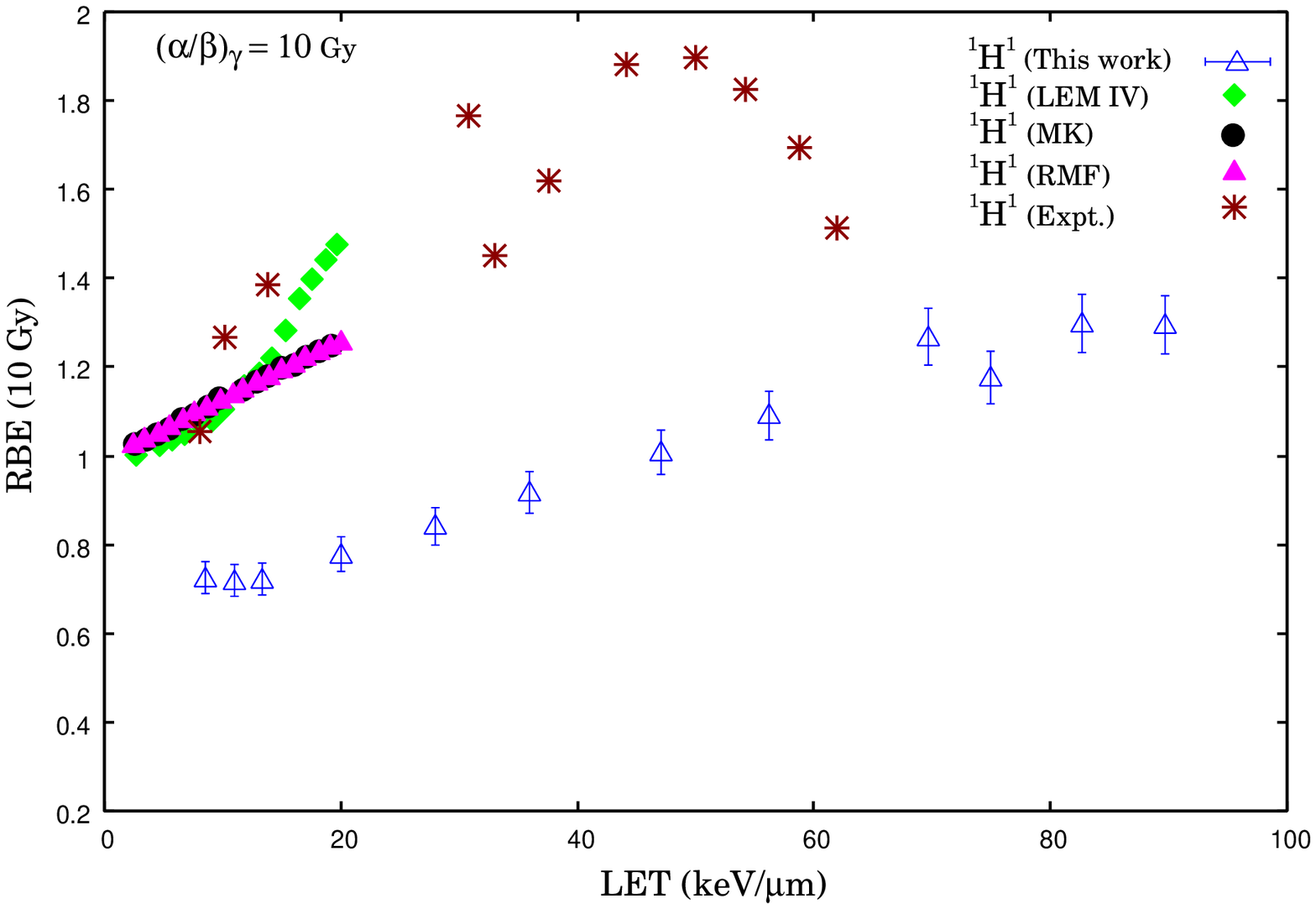}} &
\scalebox{0.45}{\includegraphics{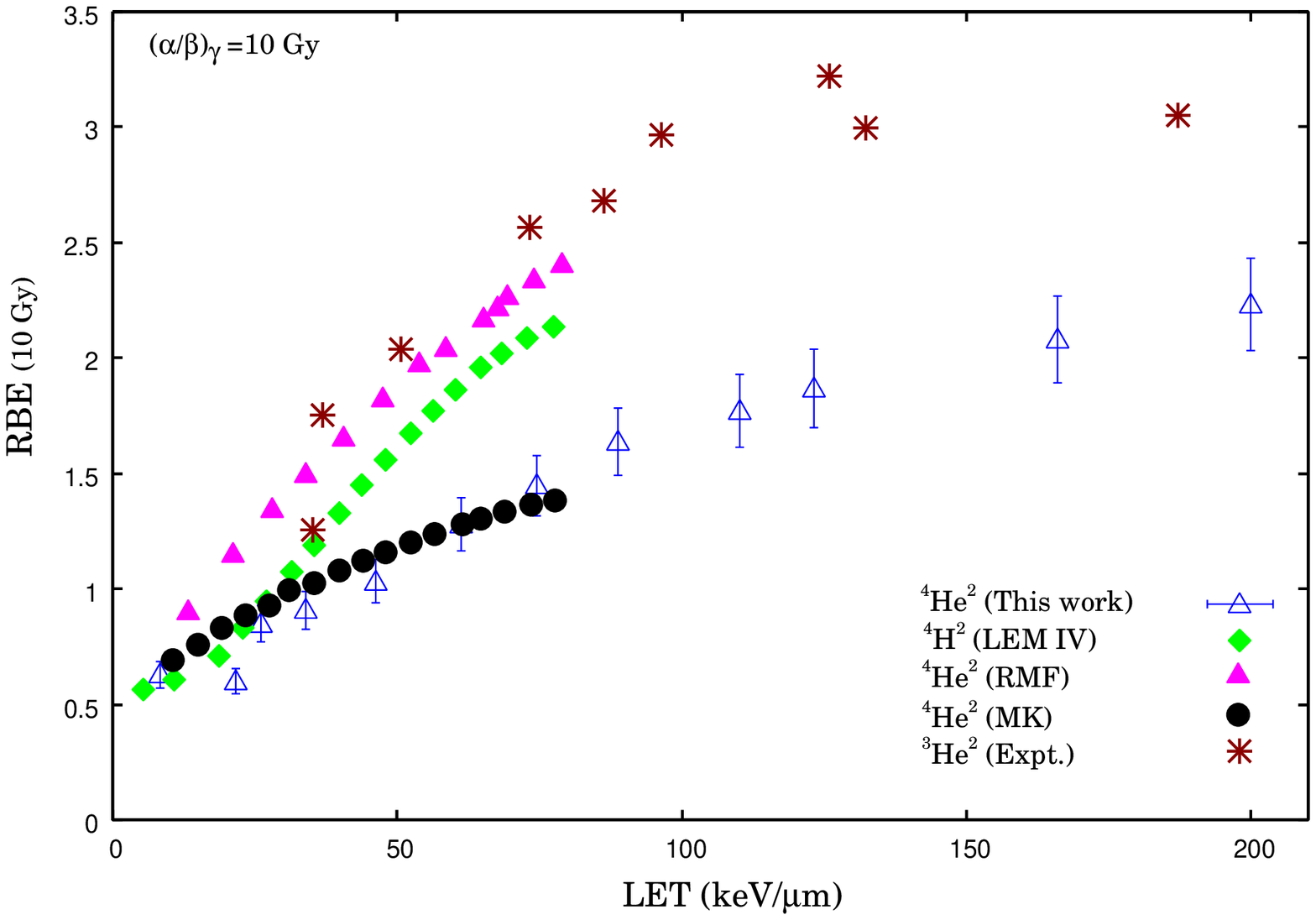}}\\
\scalebox{0.45}{\includegraphics{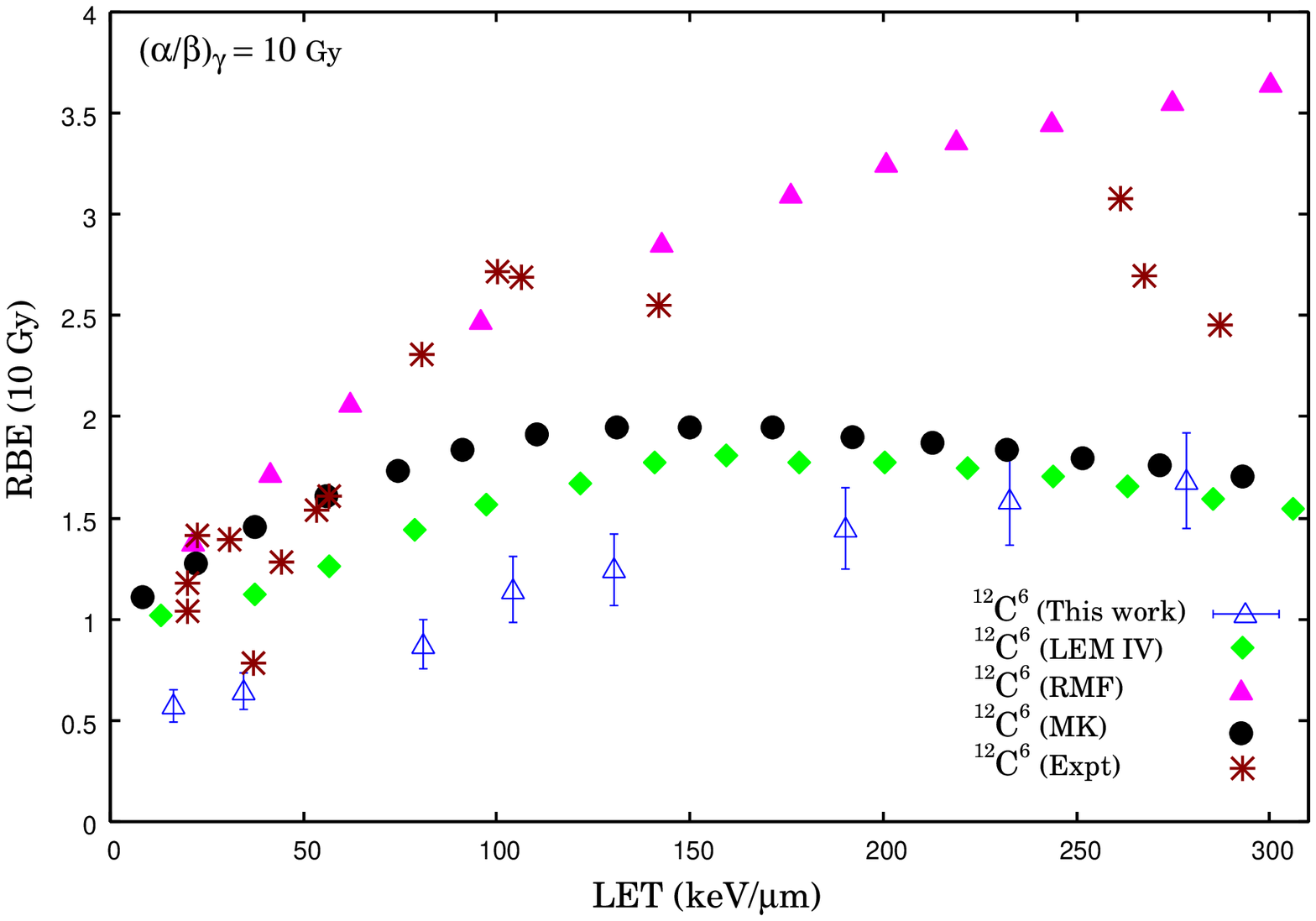}} &
\scalebox{0.60}{\includegraphics{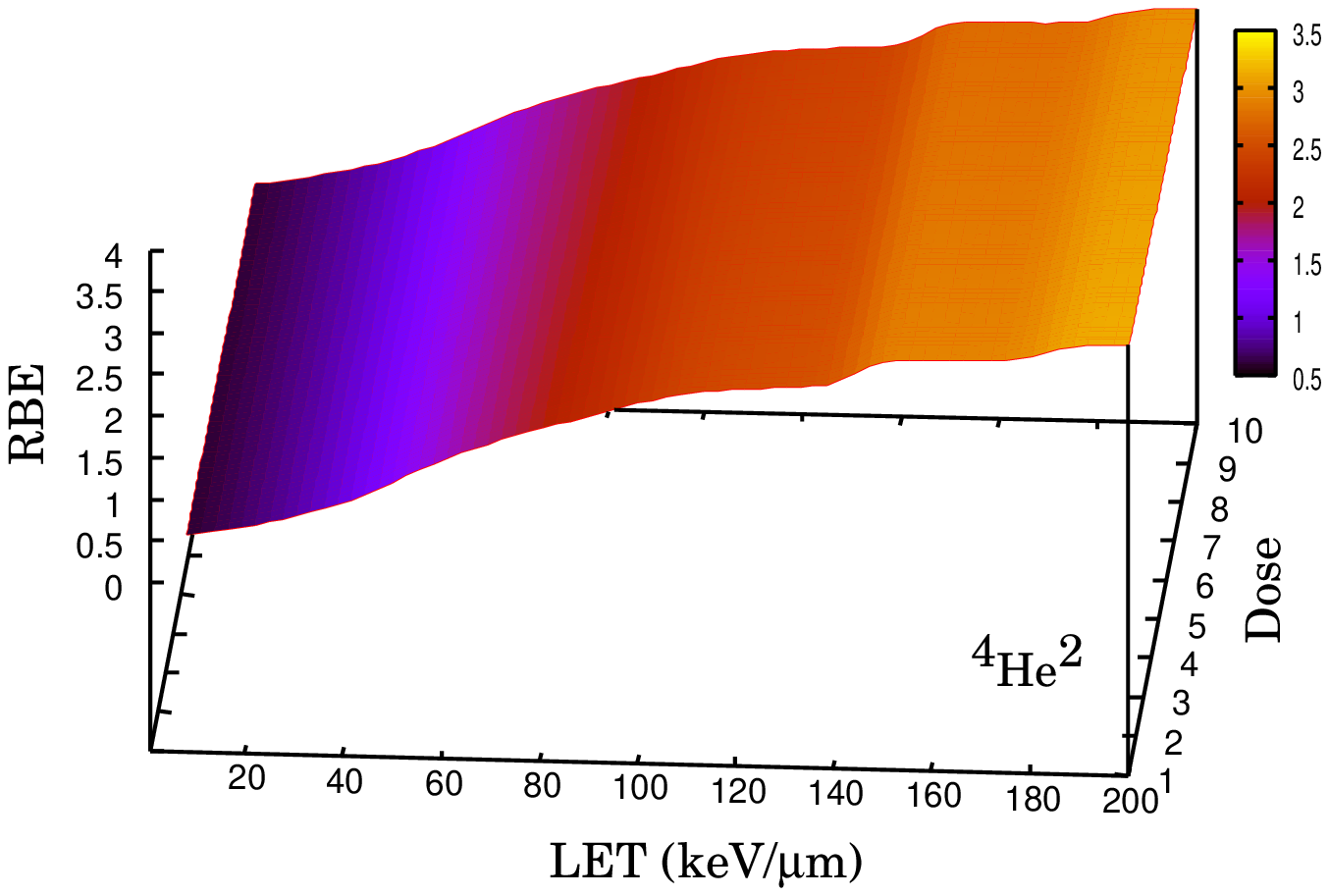}}
\end{tabular}
\caption{\label{fig3} Comparison between estimated values of RBE
and modelled values of the the LEM IV, RMF, and MK models
\cite{Stewart2018b} at $10$ Gy dose. Also, are shown the observed
RBE values from cell survival experiments at $10\%$ survival for
V79 cells \cite{Furusawa2000}.}
\end{center}
\end{figure}

To further explore the impact of LET on radio-sensitivity, we
analyse the RBE corresponding to the initial slope for different
particle types. Figure \ref{fig3} collects and compares the
predicted RBE-LET spectra, at $(\alpha /\beta )_{R} = 10$ Gy, with
the theoretical and experimental results for different radiation
species.  The proton RBE shows approximately a linear LET
dependance for LET $< 60$ keV/$\mu$m with a peak value of
approximately $1.26\pm 0.08$ at around $70$ keV/$\mu$m.  A
comparison between our estimates and the observed values shows
that in addition that the magnitude of the RBE peak value is
relatively small than the experimental value ($1.86$), the peak
appears to occur at much higher LET compared to the experimental
peak at 50 keV/$\mu$m \cite{Furusawa2000}. However, this is as
expected since the experimentally measured RBE follows a linear
relation with LET for LET less than a value that lies in the range
of $40 - 90$ keV/$\mu$m \cite{Hawkins1998}.
Another comparison of our results with the studies performed by
Stewart et al \cite{Stewart2018b} indicates different levels of
disagreement in RBE in low LET region  at threshold dose value of
$10$ Gy for protons.  The most likely reason for the disagreement
is the spatial distribution mechanism of initial DSBs
(approximated by a Poisson distribution) which has a significant
impact on the RBE for cell survival \cite{Stewart2018}.  Whereas
for low-LET ($< 50$ keV/$\mu$m) the average number of potentially
lethal DSB per track is small, and the induction of DSB can be
safely approximated by the Poisson distribution, increasing LET
causes deviation from the Poisson distribution by non-random
clustering of lethal lesions in some cells. This deprives other
cells of a lethal lesion and allows them to survive. This causes
the measured values of RBE to be lower than those indicated by the
LEM-IV and MRF models for higher LET. This structural change may
also induce an alteration in the rate of repair of lesions that
undergo transformations, thus increasing the yield of strand
breaks. Reported increase in local multiplicity of
radiation-induced strand-breaks with increasing LET could
constitute such a change in lesion structure
\cite{Ward1988,Goodhead1994} as well as a decrease in DSB yield
for LET $> 100$ keV/$\mu$m \cite{Heilmann1995}. In addition to the
shift in  maximum  of RBE, the resulting RBE is estimated to be
less than indicated by extrapolation of  the linear relationship
to higher LET values \cite{Hawkins2003,Brenner2008}.

For helium and carbon-ions, the predicted RBE shows a non-linear
behaviour for LET range explored here. The difference in the RBE
is significant in medium and high LET regions. An LET almost twice
as high is required to obtain the same RBE for helium and carbon
ions. Whereas the initial yield of DSBs is slightly higher from
helium ions than from carbon ions at the same LET in low LET
region, the biological effectiveness seems to greater for helium
ions than in carbon ions. Similar results have been reported using
low-energy light ions \cite{Folkard1996}. This can be attributed
to the difference in the track structures of the ions at the same
LET through the differences in the velocity and effective charges
of the ion. Comparison with the mechanistic models
\cite{Stewart2018b} shows that the predicted values for helium RBE
cell survival are closer to the MK model values in the medium and
large LET region. This near agreement can be attributed to the
fact that the MK model incorporates the effect of deviation from
the Poisson distribution to a compound Poisson distribution at
higher LET. For carbon ions, the deviation of the predicted values
from those of the mechanistic models is significant in low LET
region. Again, one would expect this for low-LET radiations since
a non-Poisson distribution for the induction of DSB is reduced to
a Poisson distribution of primary tracks passing through the cell
that determines the distribution of initial damage
\cite{Stewart2018}. The predicted values, however, are comparable,
within errors, with RMF and MK models in the high-LET region. We
also observed that the predicted RBE values show no significant
dose dependence for the absorbed dose $ < 3$ Gy, at low LET values
for helium ions (Fig. \ref{fig3} bottom right panel). Whether this
trend continue for reference radiation with lower $(\alpha
/\beta)_{R}$ remains to be explored.

\section{Conclusions}

Whereas, for low LET values with an average of less than one
lethal lesion per primary particle traversal, the experimental
values of RBE coincide with the RBE-LET linear relationship, the
measured RBE values are progressively lower than the predictions
of the RBE-LET linear extrapolation at higher LET values. This
decrease in the measured RBE values for high-LET radiations is an
indicator to associate the distribution of lethal lesions to
non-Poisson distributions that correlate closely with biological
effects. To account for the damage saturation correction, we
adopted the effect of a customized negative binomial distribution
of lethal lesions as a more flexible approach that accounts for
the non-random clustering and overdispersion of lethal lesions to
evaluate the variation of RBE of proton, helium and carbon ions at
higher LET. We postulated that increasing LET causes deviation
from the Poisson distribution of lethal lesions in some cells,
thereby causing a decrease in the measured values of RBE for
higher LET. We developed an equation that estimated the linear and
quadratic parameters of the model as well as the deviation of RBE
that is expected to occur because of the negative binomial
distribution of lethal lesions. The suggested deviation was
supported by observed increase $\alpha$ and $\beta$ with
increasing LET for LET below the maximum as well as by decreasing
$\alpha$ and $\beta$ for LET above the maximum.

In conclusion, the decrease in estimated RBE at high LET can be
attributed to clustering on lethal lesions in some cells causing
deviation from the Poisson distribution. The RBE results indicate
that the more sensitive cells are to radiation at low LET, lower
will be peak on RBE they attain as LET increases. Nonetheless, the
non-Poisson distribution implemented seemed to be consistent with
previously reported data for the variation of initial slope of
survival curves and RBE at high LET values of different radiation
spices. However, it is not possible to draw a definite conclusion
since the tissue dependence of RBE, which is reflected in
parameter value of reference radiation, was not explored in this
study. While assigning a high $(\alpha /\beta)_{x}$ value to the
target that actually might have a low reference parameter value
will underestimate RBE, since for low $(\alpha/\beta)_{x}$ the
average RBE could be higher. For a conclusive signature of
non-Poisson distribution of lethal lesions, work is under way to
investigate the impact of a more accurate distribution that could
result in an improved model with realistic properties overall.

\section*{Acknowledgments}
We are grateful for access to computing facility at the Kuwait
College of Science and Technology. ML is thankful to Prof. Ali
Yousef, at the Department of Mathematics, KCST,  for a number of
valuable suggestions which provided the impetus for much of this
work.

\appendix
\section{}

We use a customized negative binomial (NB) distribution for
evaluating the probability of observing $k$ lethal lesions in a
cell
\begin{equation}
P_{NB}(k) = \frac{\Gamma (k+1/r)}{\Gamma (1/r)\times k!}
\bigg(\frac{1}{1+r\mu }\bigg)^{1/r} \times
\bigg(\frac{1}{1+1/r\mu}\bigg)^{k}.
\end{equation}
The likelihood function for $N$ iid observations $(k_{1}, k_{2},
\cdots k_{N})$ is given by
\begin{displaymath}
L_{NB}=\prod_{i=1}^{N}f(k_{i})
\end{displaymath}
from which the log-likelihood function can be written as:
\begin{eqnarray}
l_{NB} & =& - \bigg(\frac{1}{r}\bigg)\times (1+k+r)\times
\ln(1+r+\mu)+ \nonumber\\
& & k\times \bigg(\ln(r)+\ln(\mu )\bigg)+ \ln \left[\Gamma\bigg(\frac{1+k+r}{r}\bigg)\right]- \nonumber\\
& & \ln[\Gamma (1+k)] -
\ln\left[\Gamma\bigg(\frac{1}{r}\bigg)\right].
\end{eqnarray}

Assuming that the correct repair distribution is NB distributed,
the probability that a break will be repaired correctly is
\begin{eqnarray}
P_{NB} & =&
\sum_{k=0}^{\infty}\frac{1}{1+k}P_{NB}(k)\nonumber\\
& = & \sum_{k=0}^{\infty} \frac{\Gamma (k+\omega )}{\Gamma (\omega
)} \times \frac{1}{1+k} p^{\omega}(1-p)^{k}\nonumber\\
 & = & \sum_{k=0}^{\infty}
 \frac{(k+\omega-1)!}{(\omega -1)!\,(k+1)!}p^{\omega}(1-p)^{k}
\end{eqnarray}
If we let $k+1=s$ and $\omega-1=z$, then
\begin{eqnarray}
P_{NB}& = & \frac{p}{z(1-p)}\sum_{s=1}^{\infty}\frac{(s+z-1)!}{s!\,(z-1)!}p^{z}(1-p)^{s}\nonumber\\
& = & \frac{pr}{(1-r)(1-p)}(1-p^{\frac{1}{r} -1}) \label{Aeqn1}
\end{eqnarray}
By fixing the average number of breaks equal to the expected
interaction rate $\mu (= \eta (\lambda_{p})n_{p})$, we can use
this to calculate the total probability of misrepair as a function
of $\lambda_{p}$. Using $p= 1/(1+r\mu)$, we expand Eq.
(\ref{Aeqn1}) to obtained
\begin{equation}
P_{NB} = \frac{1}{\mu
(1-r)}\left[1-\bigg(1+r\mu\bigg)^{1-1/r}\right].
\end{equation}


\begin{references}

\bibitem{Guerrero2002}
M. Guerrero, R. Stewart, J. Wang and X. Li, {\emph{Equivalence of
the linear�quadratic and two-lesion kinetic models}},  Phys. Med.
Biol. {\bf 47}, 3197 (2002)
\bibitem{Preston1990}
R. Preston, \emph{Mechanisms of induction of specific chromosomal
alterations}, Basic Life Sci. {\bf 53}, 329 (1990)
\bibitem{Sachs1997}
R. Sachs, P. Hahnfeld, and D. Brenner, {\emph{The link between
low-LET dose-response relations and the underlying kinetics of
damage production/repair/misrepair}},  Int. J. Radiat. Biol. {\bf
72}, 351 (1997)
\bibitem{Gerweck1994}
L. Gerweck, S. Zaidi, A. Zietman, \emph{ Multivariate determinants
of radiocurability. I: Prediction of single fraction tumor control
doses}, Int. J. Radiat. Oncol. Biol. Phys. {\bf 29}, 57 (1994)
\bibitem{Thames1985}
H. Thames, {\emph{An incomplete-repair model for survival after
fractionated and continuous irradiations}}, Int. J. Radiat. Biol.
{\bf 47}, 319 (1985)
\bibitem{Brown2003}
M. Brown, \emph{et al}., \emph{Comment on Tumor response to
radiotherapy regulated by endothelial cell apoptosis (II)},
Science, {\bf 302}, 1894 (2003)
\bibitem{Stewart2001}
R. Stewart, {\emph{Two-Lesion Kinetic Model of Double-Strand Break
Rejoining and Cell Killing}}, Radiat. Res. {\bf 56}, 365 (2001)
\bibitem{Rossi1988}
H. Rossi and M. Zaider, {\emph{in Quantitative Mathematical Models
in Radiation Biology}}, edited by J. Kiefer Springer, New York,
1988, pp. 111 - 118.
\bibitem{Brenner1990}
D. Brenner, {\emph{Track structure, lesion development, and cell
survival}}, Radiat. Res. {\bf 124}, S29 (1990)
\bibitem{Radivoy1998}
T. Radivoyevitch, D. Hoel, A. Chen, and R. Sachs,
{\emph{Misrejoining of double-strand breaks after X irradiation:
Relating moderate to very high doses by a Markov model}},  Radiat.
Res. {\bf 149}, 59 (1998)
\bibitem{Carlone2005}
M. Carlone, D. Wilkins, and P. Raaphorst, {\emph{The modified
linear-quadratic model of Guerrero and Li can be derived from a
mechanistic basis and exhibits linear-quadratic-linear
behaviour}}, Phys. Med. Biol. {\bf 50}, L9 (2005)
\bibitem{Barendsen1997}
G. Barendsen, \emph{Dose fractionation, dose rate and iso-effect
relationships for normal tissue responses}, Int. J. Radiat. Oncol.
Biol. Phys. {\bf 8}, 1981 (1982)
\bibitem{vander1985}
A. van der Kogel, \emph{Chronic effects of neutrons and charged
particles on spinal cord, lung, and rectum}, Radiat. Res. Suppl.
{\bf 8}, S208 (1985)
\bibitem{Peck1994}
J. Peck and F. Gibbs, \emph{Mechanical assay of consequential and
primary late radiation effects in murine small intestine:
alpha/beta analysis}. Radiat. Res. {\bf 138}, 272 (1994)
\bibitem{Taylor1989}
J. Taylor and D. Kim, \emph{The poor statistical properties of the
Fe-plot}, Int. J. Radiat. Biol. {\bf 56}, 161 (1989)
\bibitem{de1988}
R. de Boer, \emph{The use of the D versus dD plot to estimate the
alpha/beta ratio from iso-effect radiation damage data},
Radiother. Oncol. {\bf 11}, 361 (1988)
\bibitem{Garcia2006}
L. Garcia, J. Leblanc, D. Wilkins, and G. Raaphorst,
{\emph{Fitting the linear-quadratic model to detailed data sets
for different dose ranges}}, Phys. Med. Biol. {\bf 51}, 2813
(2006)
\bibitem{Brenner1998}
D. Brenner, \emph{et al}., \emph{The linear-quadratic model and
most other common radiobiological models result in similar
predictions of time-dose relationships}. Radiat. Res. {\bf 150},
83 (1998)
\bibitem{Curtis1986}
S.  Curtis, \emph{Lethal and potentially lethal lesions induced by
radiation - a unified repair model}. Radiat. Res. {\bf 106}, 252
(1986)
\bibitem{Hawkins1996}
R. Hawkins, \emph{ A microdosimetric-kinetic model of cell death
from exposure to ionizing radiation of any LET, with experimental
and clinical applications}. Int. J. Radiat. Biol. {\bf 69}, 739
(1996)
\bibitem{Obaturov1993}
G. Obaturov, V. Moiseenko, A. Filimonov, \emph{Model of mammalian
cell reproductive death. I. Basic assumptions and general
equations}. Radiat. Environ. Biophys. {\bf 32}, 285 (1993)
\bibitem{Tobias1995}
C. Tobias, \emph{The repair-misrepair model in radiobiology:
comparison to other models}, Radiat. Res. Suppl. {\bf 8}, S77
(1985)
\bibitem{Hawkins1998}
R.  Hawkins, \emph{A microdosimetric-kinetic theory of the
dependence of the RBE for cell death on LET}, Med. Phys. {\bf 25},
1157 (1998)
\bibitem{Zaider1998}
M. Zaider, \emph{There is no mechanistic basis for the use of the
linear-quadratic expression in cellular survival analysis}, Med.
Phys. {\bf 25}, 791 (1998)
\bibitem{Hawkins2017}
R. Hawkins, {\emph{Effect of heterogeneous radiosensitivity on the
survival, alpha beta ratio and biologic effective dose calculation
of irradiated mammalian cell populations}},  Clin. Transl. Radiat.
Oncol. {\bf 4}, 32 (2017)
\bibitem{Sachs1998}
Sachs RK and D.  Brenner, {\emph{The mechanistic basis of the
linear-quadratic formalism}}, Med. Phys. {\bf 25}, 2071 (1998)
\bibitem{Brenner2008}
D. Brenner, {\emph{The linear-quadratic model is an appropriate
methodology for determining isoeffective doses at large doses per
fraction}}, Semin. Radiat. Oncol. {\bf 18}, 234  (2008)
\bibitem{Kirk2009}
J. Kirkpatrick, D. Brenner and C. Orton, {\emph{The
linear-quadratic model is inappropriate to model high dose per
fraction effects in radiosurgery}},  Med. Phys. {\bf 36}, 3381
(2009)
\bibitem{Hawkins2003}
R. Hawkins,  {\emph{A Microdosimetric-Kinetic model for the effect
of non-Poisson distribution of lethal lesions on the variation of
RBE with LET}}, Radiat. Res. {\bf 160}, 61  (2003)
\bibitem{Harrison2014}
X. Harrison, {\emph{Using observation-level random effects to
model overdispersion in count data in ecology and evolution}},
PeerJ, {\bf 2}, e616 (2014)
\bibitem{Iliakins2015}
G. Iliakis, T. Murmann and A. Soni, {\emph{Alternative end-joining
repair pathways are the ultimate backup for abrogated classical
non-homologous end-joining and homologous recombination repair:
implications for the formation of chromosome translocations}},
Mutat. Res. Toxicol. Environ Mutagen, {\bf 793}, 166 (2015)
\bibitem{Shelke2015}
S. Shelke and B. Das, {\emph{Dose response and adaptive response
of non-homologous end joining repair genes and proteins in resting
human peripheral blood mononuclear cells exposed to � radiation}},
Mutagenesis, {\bf 30}, 365 (2015)
\bibitem{Hawkins2017b}
R. Hawkins, {\emph{Biophysical Models, Microdosimetry and the
Linear Quadratic Survival Relation}}, Ann. Radiat. Ther. Oncol.
{\bf 1}, 1013 (2017)
\bibitem{Ward1988}
J.  Ward, {\emph{DNA Damage Produced by Ionizing Radiation in
Mammalian Cells: Identities, Mechanisms of Formation, and
ReparabilityProg}}, Nucleic Acid Res. Mol. Biol. {\bf 35}, 95
(1988)
\bibitem{Goodhead1994}
D. Goodhead, {\emph{Initial events in the cellular effects of
ionizing radiations: Clustered damage in DNA}}, Int. J. Radiat.
Biol. {\bf 65}, 7 (1994).
\bibitem{Heilmann1995}
J. Heilmann, G. Taucher-Scholz and G. Kraft, {\emph{Induction of
DNA double-strand breaks in CHO-K1 cells by carbon ions}}, Int. J.
Radiat. Biol. {\bf 68}, 153 (1995)
\bibitem{Nowak2000} E.
Gudowska-Nowak, M. Kramer,  G. Kraft  and G. Taucher-Scholz,
{\emph{Compound Poisson Statistics and Models of Clustering of
Radiation Induced DNA Double Strand Breaks}}.
arXiv:physics/0011071v1 [physics.bio-ph]  (2000)
\bibitem{Virsik1981}
R. Virsik and D. Harder, {\emph{Statistical
interpretation of the overdispersed distribution of
radiation-induced dicentric chromosome aberrations at high LET}},
Radiat. Res. {\bf 85}, 13 (1981)
\bibitem{Goodwin1994}
E.  Goodwin, E. Blakely and C. Tobias, {\emph{ Chromosomal damage
and repair in G1-phase Chinese hamster ovary cells exposed to
charged-particle beams}}. Radiat. Res. {\bf 138}, 343 (1994)
\bibitem{Elsasser2010}
T. Elsasser, \emph{et al}., {\emph{Quantification of the relative
biological effectiveness for ion beam radiotherapy: direct
experimental comparison of proton and carbon ion beams and a novel
approach for treatment planning}}, Int. J. Radiat. Oncol. Biol.
Phys. {\bf 78}, 1177 (2010)
\bibitem{Friedrich2012}
T. Friedrich T, U. Scholz, T. Elsasser, M. Durante M and M.
Scholz, {\emph{Calculation of the biological effects of ion beams
based on the microscopic spatial damage distribution pattern}},
Int. J. Radiat. Biol. {\bf 88}, 103 (2012)
\bibitem{Hawkins1994}
R. Hawkins, {\emph{ A statistical theory of cell killing by
radiation of varying linear energy transfer}}. Radiat. Res. {\bf
140}, 366 (1994)
\bibitem{Inaniwa2010}
T. Inaniwa, \emph{ et al}., {\emph{Treatment planning for a
scanned carbon beam with a modified microdosimetric kinetic
model}},  Phys. Med. Biol. {\bf 55}, 6721 (2010)
\bibitem{Carlson2008}
D. Carlson, R.  Stewart, V. Semenenko and G. Sandison, {\emph{
Combined use of Monte Carlo DNA damage simulations and
deterministic repair models to examine putative mechanisms of cell
killing}},  Radiat. Res. {\bf 169}, 447 (2008)
\bibitem{Frese2012}
M. Frese, V. Yu, R. Stewart and D.  Carlson, {\emph{A
mechanism-based approach to predict the relative biological
effectiveness of protons and carbon ions in radiation therapy}},
Int. J. Radiat. Oncol. Biol. Phys. {\bf 83}, 442 (2012)
\bibitem{Streitmatter2017}
S. Streitmatter, R. Stewart, P. Jenkins and T. Jevremovic T,
{\emph{DNA double strand break (DSB) induction and cell survival
in iodine-enhanced computed tomography (CT)}}, Phys. Med. Biol.
{\bf 62}, 6164 (2017)
\bibitem{Wang2018}
W. Wang, \emph{et al}., {\emph{Modelling of cellular survival
following radiation-induced DNA double-strand breaks}}, Sci. Rep.
{\bf 8}, 16202 (2018)
\bibitem{Karge2017}
C. Karge and P. Peschke, {\emph{RBE and related modeling in
carbon-ion therapy}}, Phys. Med. Biol. {\bf 63}, 01TR02 (2017).
\bibitem{McMahon2016}
S. McMahon, J. Schuemann, H. Paganetti and K. Prise,
{\emph{Mechanistic modelling of DNA repair and cellular survival
following radiation-induced DNA damage}}, Sci. Rep. {\bf 6}, 33290
(2016)
\bibitem{MCDScode}
V. Semenenko and R. Stewart, {\emph{Fast Monte Carlo Simulation of
DNA Damage Formed by Electrons and Light Ions Phys}}, Med. Biol.
{\bf 51}, 1693 (2006)
\bibitem{Stewart2015}
R. Stewart RD, \emph{et al}., {\emph{ Rapid MCNP simulation of DNA
double strand break (DSB) relative biological effectiveness (RBE)
for photons, neutrons, and light ions}}, Phys. Med. Biol. {\bf
60}, 8249 (2015)
\bibitem{Stewart2018}
R. Stewart, {\emph{Induction of DNA damage by light ions relative
to 60Co gamma-rays}}, Int. J. Part. Ther. {\bf 5}, 25 (2018)
\bibitem{Prise1990}
K. Prise KM, M. Folkard M, S. Davies and B. Michael, \emph{The
irradiation of V79 mammalian cells by protons with energies below
2 MeV. Part II. Measurement of oxygen enhancement ratios and DNA
damage}. Int J Radiat Biol. {\bf 58}, 261 (1990)
\bibitem{Friedrich2013}
T. Friedrich \emph{eta al}., {\emph{Systematic analysis of RBE and
related quantities using a database of cell survival experiments
with ion beam irradiation}},  J. Radiat. Res. {\bf 54}, 494,
(2013)
\bibitem{Paganetti2002}
H. Paganetti \emph{et al}.,  {\emph{Relative biological
effectiveness (RBE) values for proton beam therapy}}, Int. J
Radiat. Oncol. Biol. Phys. {\bf 53}, 407 (2002)
\bibitem{Paganetti2014}
H. Paganetti \emph{et al}., {\emph{Relative biological
effectiveness (RBE) values for proton beam therapy. Variation as a
function of biological endpoints, dose, and linear energy
transfer}}, Phys. Med. Biol. {\bf 59}, R419 (2006)
\bibitem{Giovannini2016}
G. Giovannini \emph{et al}., {\emph{Variable RBE in proton
therapy: Comparison of different model predictions and their
influence on clinical-like scenarios }}, Radiat. Oncol. {\bf 11},
68 (2016)
\bibitem{Leeuwen2018}
C. van Leeuwen \emph{et al}., {\emph{The alfa and beta of tumours:
A review of parameters of the linear-quadratic model, derived from
clinical radiotherapy studies }}, Radiat. Oncol. {\bf 13}, 96
(2018)
\bibitem{Mario2017}
M. Carante and F. Ballarini, {\emph{Modelling cell death for
cancer hadrontherapy}}, AIMS Biophysics, {\bf 4}, 465 (2017)
\bibitem{Furusawa2000}
Y. Furusawa, \emph{et al}., {\emph{Inactivation of aerobic and
hypoxic cells from three different cell lines by accelerated 3He-,
12C- and 20Ne-Ion beams}}, Radiat. Res. {\bf 154}, 485 (2000)
\bibitem{Stewart2018b}
R. Stewart, \emph{et al}., {\emph{A comparison of
mechanism-inspired models for particle relative biological
effectiveness (RBE)}}, Med. Phys. {\bf 45}, e928 (2018)
\bibitem{Butkus2018}
M. Butkus, R. Stewart, Z. Chen Z and D. Carlson, {\emph{Double
strand break (DSB) complexity and proximity effects within the
repair-misrepair fixation (RMF) model for improved predictions of
cell survival from heavy ions}}, Med. Phys. {\bf 45 }, e532 (2018)
\bibitem{Goodhead1994b}
D. Goodhead, {\emph{Initial events in the cellular effects of
ionizing radiations: clustered damage in DNA}},  Int J. Radiat.
Biol. {\bf 65}, 7 (1994)
\bibitem{Hada2008}
M. Hada, A. Georgakilas, {\emph{Formation of clustered DNA damage
after high-LET irradiation: a review}},  J. Radiat. Res. {\bf 49},
203 (2008)
\bibitem{Goodhead2009}
D. Goodhead, {\emph{Fifth Warren K. Sinclair keynote address:
issues in quantifying the effects of low-level radiation}}, Health
Phys. {\bf 97}, 394 (2009)
\bibitem{David2014}
A. Davis and D. Chen, {\emph{Complex DSBs: a need for resection}},
Cell Cycle, {\bf 13}, 3796 (2014)
\bibitem{Hunnagl2015}
A. Hufnagl, \emph{et al}., {\emph{The link between cell cycle
dependent radiosensitivity and repair pathways: a model based on
the local, sister-chromatid conformation dependent switch between
NHEJ and HR}}, DNA Repair, {\bf 27}, 28 (2015)
\bibitem{Folkard1996}
M. Folkard, \emph{et al}., {\emph{Inactivation of V79 cells by
low-energy protons, deuterons and helium-3 ions}}, Int. J. Radiat.
Biol. {\bf 69}, 796 (1996)

\end{references}
\end{document}